\documentclass[a4paper,11pt]{article}

\usepackage{jinstpub} 

\usepackage{epsfig}
\usepackage{epstopdf}
\usepackage{color} 
\usepackage{url} 

\usepackage{lineno}

\title{Test of $^{116}$CdWO$_4$ and Li$_2$MoO$_4$ scintillating bolometers in the CROSS underground facility with upgraded detector suspension}

\author[a]{A.~Ahmine,}
\author[b]{I.C.~Bandac,}
\author[c]{A.S.~Barabash,}
\author[d]{V.~Berest,}
\author[e]{L.~Berg{\'e},}
\author[b,f,g]{J.M.~Calvo-Mozota,}
\author[h,i]{P.~Carniti,}
\author[e]{M.~Chapellier,}
\author[j]{I.~Dafinei,}
\author[k,l]{F.A.~Danevich,}
\author[e]{T.~Dixon,}
\author[e]{L.~Dumoulin,}
\author[d]{F.~Ferri,}
\author[e]{A.~Giuliani,}
\author[h]{C.~Gotti,}
\author[d]{P.~Gras,}
\author[d,*]{D.L.~Helis,}
\author[m]{A.~Ianni,}
\author[e]{L.~Imbert,}
\author[d]{H.~Khalife,}
\author[k]{V.V.~Kobychev,}
\author[c]{S.I.~Konovalov,}
\author[e]{P.~Loaiza,}
\author[e]{P.~de~Marcillac,}
\author[e]{S.~Marnieros,}
\author[e]{C.A.~Marrache-Kikuchi,}
\author[n,o]{M.~Martinez,}
\author[d]{C.~Nones,}
\author[e]{E.~Olivieri,}
\author[n]{A.~Ortiz de Sol\'orzano,}
\author[e]{Y.~Peinaud,}
\author[h]{G.~Pessina,}
\author[e]{D.V.~Poda,}
\author[e]{Th.~Redon,}
\author[e]{Ph.~Rosier,}
\author[e]{J.A.~Scarpaci,}
\author[k,m]{V.I.~Tretyak,}
\author[c]{V.I.~Umatov,}
\author[a]{M.~Velazquez,}
\author[k]{M.M.~Zarytskyy,}
\author[d]{and A.~Zolotarova}

\affiliation[a]{Univ. Grenoble Alpes, CNRS, Grenoble INP (Institute of Engineering Univ. Grenoble Alpes), SIMAP, 38000 Grenoble, France}
\affiliation[b]{Laboratorio Subterr\'aneo de Canfranc, 22880 Canfranc-Estaci\'on, Spain}
\affiliation[c]{National Research Center Kurchatov Institute, Kurchatov Complex of Theoretical and Experimental Physics, 117218 Moscow, Russia}
\affiliation[d]{IRFU, CEA, Université Paris-Saclay, 91191 Saclay, France}
\affiliation[e]{Universit\'e Paris-Saclay, CNRS/IN2P3, IJCLab, 91405 Orsay, France}
\affiliation[f]{Escuela Superior de Ingenier\'ia, Ciencia y Tecnolog\'ia, Universidad Internacional de Valencia -- VIU, 46002 Valencia, Spain}
\affiliation[g]{Escuela Superior de Ingenier\'ia y Tecnolog\'ia, Universidad Internacional de La Rioja, 26006 Logro\~no, Spain}
\affiliation[h]{INFN Sezione di Milano-Bicocca, I-20126 Milan, Italy}
\affiliation[i]{Universit\`{a} di Milano-Bicocca, Milano I-20126, Italy} 
\affiliation[j]{INFN Sezione di Roma, I-00185 Rome, Italy}
\affiliation[k]{Institute for Nuclear Research of NASU, 03028 Kyiv, Ukraine}
\affiliation[l]{INFN Sezione di Roma Tor Vergata, I-00133 Rome, Italy}
\affiliation[m]{INFN Laboratori Nazionali del Gran Sasso, I-67100 Assergi (AQ), Italy}
\affiliation[n]{Centro de Astropart\'iculas y F\'isica de Altas Energ\'ias, Universidad de Zaragoza, 50009 Zaragoza, Spain}
\affiliation[o]{ARAID Fundaci\'on Agencia Aragonesa para la Investigaci\'on y el Desarrollo, 50018  Zaragoza, Spain}
\affiliation[*]{Currently at INFN Laboratori Nazionali del Gran Sasso (Assergi) and Gran Sasso Science Institute (L'Aquila), Italy}

\emailAdd{denys.poda@ijclab.in2p3.fr}

\abstract{
In preparation to the CROSS double-beta decay experiment, we installed a new detector suspension with magnetic dumping inside a pulse-tube cryostat of a dedicated low-background facility at the Canfranc underground laboratory (Spain). The suspension was tested with two scintillating bolometers based on large-volume $^{116}$CdWO$_4$ and Li$_2$MoO$_4$ crystals. The former, already tested detector, was used as a reference device for testing new noise conditions and for comparing bolometric performance of an advanced Li$_2$MoO$_4$ crystal developed in the framework of the CLYMENE project, in view of next-generation double-beta decay experiments like CUPID. We cooled down detectors to 15 mK and achieved high performance for all tested devices. In particular both $^{116}$CdWO$_4$ and Li$_2$MoO$_4$ bolometers demonstrated the energy resolution of $\sim$6 keV FWHM for the 2.6 MeV $\gamma$ quanta, among the best for thermal detectors based on such compounds. The baseline noise resolution (FWHM) of the $^{116}$CdWO$_4$ detector was improved by $\sim$2 keV, compared to the best previous measurement of this detector in the CROSS facility, while the noise of the Ge-based optical bolometer was improved by a factor 2, to $\sim$100 eV FWHM. Despite of the evident progress in the improving of noise conditions of the set-up, we see high-frequency harmonics of a pulse-tube induced noise, suggesting a noise pick-up by cabling. Another Ge light detector was assisted with the signal amplification exploiting the Neganov-Trofimov-Luke effect, which allowed to reach $\sim$20 eV FWHM noise resolution by applying 60 V electrode bias. A highly-efficient particle identification was achieved with both detectors, despite a low scintillation efficiency of the Li$_2$MoO$_4$ material. 
The radiopurity level of the Li$_2$MoO$_4$ crystal is rather high; only traces of $^{210}$Po and $^{226}$Ra were detected ($\sim$0.1 mBq/kg each), while the $^{228}$Th activity is expected to be at least an order of magnitude lower, as well as a $^{40}$K activity is found to be < 6 mBq/kg. 
}


\begin{document}
\maketitle
\flushbottom

\section{Introduction}
\label{sec:intro}

Low-temperature detectors \cite{Enss:2005a,Enss:2005,Enss:2008} became more widely used in searches for rare and/or forbidden in the Standard Model (SM) processes, in particularly double-beta ($\beta\beta$) decay \cite{Pirro:2017,Biassoni:2020,Poda:2021,Zolotarova:2021a}. The SM process, two-neutrino ($2\nu$) $\beta\beta$ decay, is the rarest nuclear disintegration ever observed; to date this process is detected with the half-lives $T_{1/2} \sim 10^{19}$--$10^{24}$ yr in about dozen nuclei \cite{Pritychenko:2023,Barabash:2020,Belli:2020a,Saakyan:2013} from the full list of $\sim$70 natural potentially $\beta\beta$-active isotopes \cite{Tretyak:2002}. A virtual exchange of Majorana neutrinos with a finite mass in the $\beta\beta$ decay, would result to emission of two electrons only, neutrino-less ($0\nu$) $\beta\beta$ decay, and therefore to violation of the total lepton number, which requires existence of physics beyond the SM (see recent reviews \cite{Agostini:2022,Workman:2022,Dolinski:2019} and references there). Presently, no evidence of the $0\nu\beta\beta$ decay process has been reported and the most sensitive experiments set lower limits on the half-lives at the level of $T_{1/2} > 10^{24}$--$10^{26}$ yr \cite{Arnold:2015,Anton:2019,Adams:2020,Agostini:2020,Azzolini:2022,Augier:2022,Arnquist:2023,Abe:2023}. In order to achieve a further progress in $0\nu\beta\beta$ decay searches, new-generation experiments are going to use detector technologies capable of a long-term ($\sim$10 yr) operation of a large mass (tonne-scale) of the $\beta\beta$ isotope of interest embedded in the detector material, and characterized by as high as possible energy resolution and ultra-low (ideally zero) background \cite{Agostini:2022,Adams:2022b,DAndrea:2021,Biassoni:2020,Shimizu:2019,GomezCadenas:2019}.

Scintillating low-temperature detectors with the simultaneous readout of the crystals heat and scintillation (so-called scintillating bolometers) represent a great interest for high-sensitivity $\beta\beta$ decay experiments thanks to the different choice of materials, typically high detector performance at low temperatures, the scalability to a tonne-scale, and the scintillation assisted particle identification, a key importance tool of an active background rejection \cite{Pirro:2005ar,Pirro:2017,Poda:2017,Bellini:2018,Biassoni:2020,Poda:2021,Zolotarova:2021a}. 
In particular, two next-generation experiments CUPID \cite{Wang:2015raa,CUPIDInterestGroup:2019inu} and AMoRE \cite{Alenkov:2015} will exploit this technique to investigate $\beta\beta$ decay of $^{100}$Mo (with the decay energy $Q_{\beta\beta}$  = 3034 keV \cite{Wang:2021a}). It is worth noting that several other $\beta\beta$ isotopes, like $^{82}$Se (2998 keV), $^{116}$Cd (2813 keV), and $^{130}$Te (2528 keV) also represent interest for CUPID \cite{Wang:2015raa} and/or other possible multi-isotope experiments \cite{Giuliani:2018,Ettengruber:2022,Agostini:2023}.

Lithium molybdate (Li$_2$MoO$_4$) crystal scintillator \cite{Barinova:2010,Cardani:2013} is now a primary choice for CUPID \cite{CUPIDInterestGroup:2019inu}, motivated by the results of the LUMINEU R\&D on scintillating bolometers based on lithium molybdate crystals produced from molybdenum enriched in $^{100}$Mo (Li$_2$$^{100}$MoO$_4$) \cite{Armengaud:2017,Grigorieva:2017,Poda:2017a}, reinforced by a small-scale demonstrator CUPID-Mo \cite{Armengaud:2020a,Augier:2022}. A cubic crystal with a 45 mm side is a target size of the CUPID \cite{Armatol:2021a,Armatol:2021b,Alfonso:2022}. The samples used in the CUPID R\&D program, have been taken from the batch of 32 crystals produced, following the LUMINEU protocol, for the CROSS experiment \cite{Bandac:2020}. The CROSS project is devoted to the development of a technology of metal-coated bolometers capable of particle identification of near surface interactions \cite{Bandac:2020,Khalife:2020,Khalife:2020a,Zolotarova:2020,Bandac:2021,Khalife:2021}, as a potential tool for further background reduction in next-generation experiments, particularly in CUPID-Reach (a factor 5 improved background compared to the baseline version) \cite{CUPIDInterestGroup:2019inu}. 
Moreover, Li$_2$MoO$_4$ was chosen by the AMoRE collaboration \cite{Kim:2020a,Kim:2020b,Kim:2022a,Kim:2022b,Kim:2023}, aiming at searching for $^{100}$Mo double-beta decay with metallic magnetic calorimeters.

In view of the needs of a large-scale production of Li$_2$$^{100}$MoO$_4$ crystals for CUPID, a project CLYMENE \cite{Velazquez:2017} has been funded in France for the development of Li$_2$MoO$_4$ crystals using Czochralski (Cz) growth technique, specifically developed for bulk Li$_2$MoO$_4$ crystals by means of modelling and numerical simulations of the growth process. A progress in the growth of large-volume  Li$_2$MoO$_4$ crystals has been recently achieved by the CLYMENE collaboration \cite{Velazquez:2017,Stelian:2018,Stelian:2020,Ahmine:2022a,Ahmine:2022b}. A first test of the advanced CLYMENE crystal as a scintillating bolometer is a subject of the present paper. In order to make a comparison of the Li$_2$MoO$_4$ detector performance, we operated simultaneously another scintillating bolometer based on cadmium tungstate produced from $^{116}$Cd-enriched cadmium. The later detector was used as a reference for these measurements, realized in the Canfranc underground laboratory (LSC, Spain) using the CROSS cryogenic facility with upgraded detector suspension.

\section{Crystals and low-temperature detectors}

\subsection{Crystal scintillators}

\subsubsection{Cadmium tungstate enriched in $^{116}$Cd}
\label{sec:CWO_crystal}

For this study we used a cadmium tungstate crystal, which was produced from cadmium enriched in $^{116}$Cd ($\sim$82\%) \cite{Barabash:2011} and utilized in the Aurora $\beta\beta$ experiment with a two-crystal scintillation detector at the Gran Sasso underground laboratory (Italy) \cite{Barabash:2018}. The 1.9~kg ingot of the $^{116}$CdWO$_4$ crystal scintillator with high optical quality was grown using a purified enriched material and the low-temperature gradient Cz growth technique. The crystal boule was cut in five samples, as illustrated in \cite{Barabash:2016a}, and three of them have been already tested as scintillating bolometers demonstrating high performance \cite{Barabash:2016,Helis:2020,Helis:2021}. For this study, we used the crystal No. 2 ($\oslash$45$\times$46.7~mm, 582~g), as defined in \cite{Barabash:2011,Barabash:2016a,Barabash:2018}, cut from the bottom part of the ingot. The presence of $^{113}$Cd (2.1\% in the enriched material) and $^{113m}$Cd are responsible for the dominant radioactivity (0.5~Bq/kg in total) in the crystal bulk \cite{Barabash:2011}. Thanks to the applied purification methods, the contamination by radionuclides from U/Th chain is at significantly lower level \cite{Barabash:2011,Barabash:2018}: the total $\alpha$ activity is a few mBq/kg (the dominant sources are $^{238}$U, $^{234}$U, and $^{210}$Po with a similar activity of $\sim$0.6~mBq/kg, while the $^{228}$Th activity is only tens $\mu$Bq/kg).

\subsubsection{Lithium molybdate (CLYMENE project)}
\label{sec:LMO_crystal}

Another crystal scintillator studied in the present work is Li$_2$MoO$_4$, which was developed within the R\&D program of the CLYMENE project \cite{Velazquez:2017,Stelian:2018,Stelian:2020,Ahmine:2022a,Ahmine:2022b}.  The optimization of the Cz growth process was guided by COMSOL multiphysics modelling and numerical simulations in conjunction with well chosen experimental validations \cite{Ahmine:2022a}. The furnace configuration and the crystal growth parameters were optimized to produce large-volume ingots with a regular shape (5 cm in diameter, about 15 cm in length) and with reduced thermal stresses \cite{Stelian:2018,Stelian:2020,Ahmine:2022b}, in contrast to the first large CLYMENE crystal (0.23~kg) which exhibited a partial cleavage starting from the bottom part of the ingot \cite{Velazquez:2017}. Two parts of this crystal with \cite{Velazquez:2017} and without \cite{Buse:2018} a cleavage were tested as low-temperature scintillating detectors, particularly showing the impact of the crystal crack on the phonon propagation. The first large-mass (0.82~kg) ingot grown in the optimized Cz furnace \cite{Ahmine:2022a} exhibits regular shape (about 15 cm of the cylindrical part), good structural quality and no mechanical cleavage upon cooling, processing and shaping procedures. We used the bottom part of the ingot to produce a Li$_2$MoO$_4$ sample ($\oslash$48$\times$44~mm, 245~g) for this work.

\subsection{Construction of low-temperature detectors}
\label{sec:Detectors}

\subsubsection{$^{116}$CdWO$_4$ scintillating bolometer}

\begin{figure}
\centering
\includegraphics[width=1.0\textwidth]{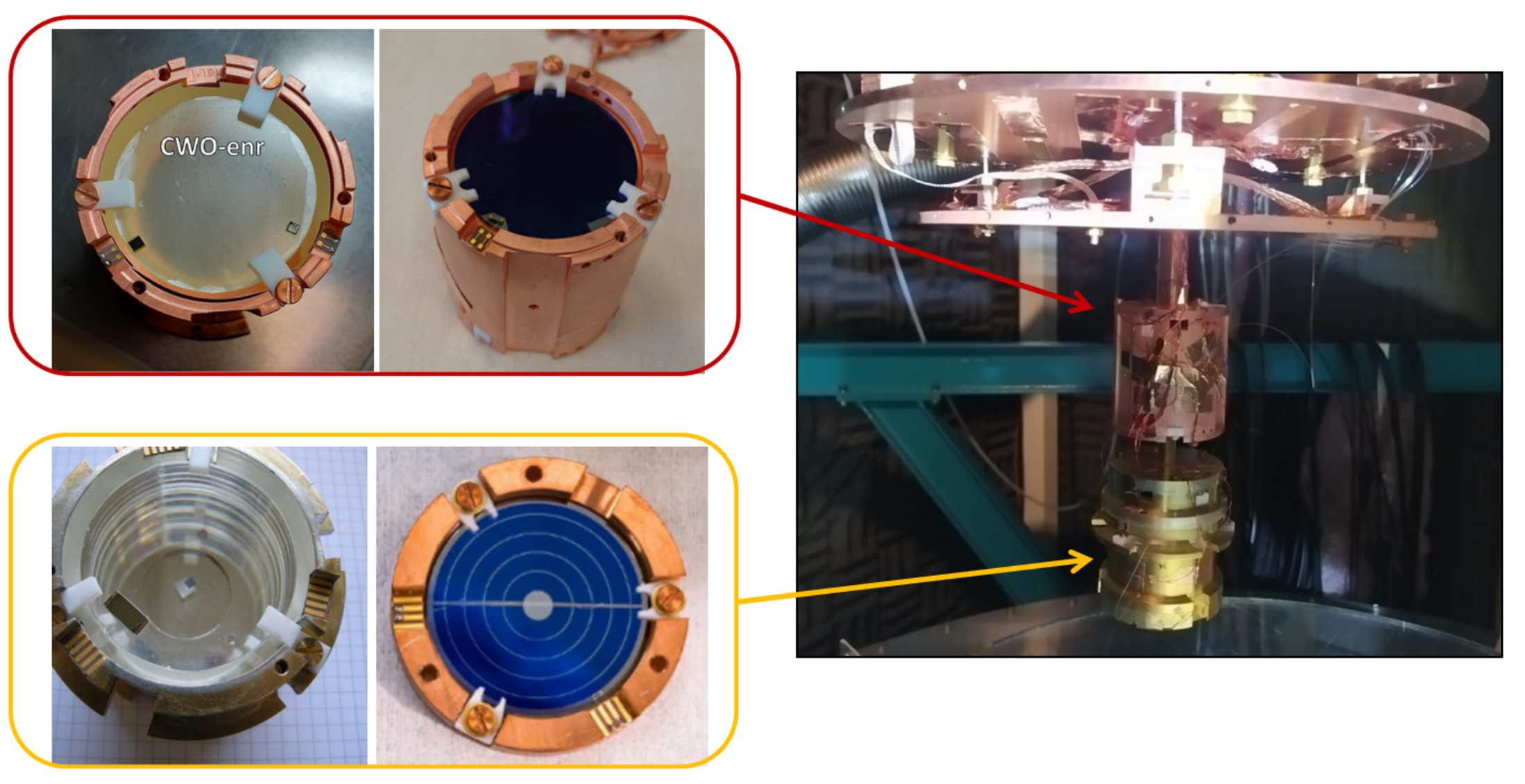}
\caption{Scintillating bolometers based on $^{116}$CdWO$_4$ (Left, top) and Li$_2$MoO$_4$ (Left, bottom) crystal scintillators. Each module has a bolometric Ge light detector. One Ge device has concentric Al electrodes to exploit signal amplification based on the Neganov-Trofimov-Luke effect. Both modules were connected to the detector plate of the CROSS cryogenic facility (Right).}
\label{fig:Detectors}
\end{figure}

A $^{116}$CdWO$_4$ scintillating bolometer was developed using the crystal sample described in Sec. \ref{sec:CWO_crystal} and following the CUPID-Mo single module design \cite{Armengaud:2020a}. The crystal was held in the Cu housing with the help of PTFE pieces and a Ge high purity wafer ($\oslash$44$\times$0.18 mm) used for photodetection was placed from one side of the crystal, as illustrated in figure \ref{fig:Detectors} (Left, top). The Ge slab was coated from both sides with a $\sim70$~nm layer of SiO, aiming at reducing the reflection of scintillation photons \cite{Mancuso:2014,Azzolini:2018tum,Armengaud:2020a}. The light collection inside the Cu holder was enhanced by a reflective film (Vikuiti{\texttrademark}) surrounding the lateral side of the crystal scintillator.

\nopagebreak

The $^{116}$CdWO$_4$ crystal and the Ge wafer were equipped with Neutron Transmutation Doped Ge \cite{Haller:1994} thermistors (NTDs), coupled with Araldite\textregistered  Rapid epoxy. The $^{116}$CdWO$_4$ crystal was instrumented with a larger NTD (3$\times$3$\times$1~mm), while a smaller sensor (1$\times$3$\times$1~mm) was used for the Ge wafer. A heating element made of P-doped Si \cite{Andreotti:2012} was also glued on the $^{116}$CdWO$_4$ crystal and used for the detector optimization (study of NTD working points) and for the stabilization of the detector thermal gain \cite{Alessandrello:1998}, which can be affected by even tiny ($\sim$10~$\mu$K) temperature variations in time which are typical for dilution refrigerators.  

The $^{116}$CdWO$_4$ scintillating bolometer has been already tested in the CROSS set-up, during its commissioning \cite{Helis:2020} and in the first scientific run \cite{Helis:2021} (results of another detectors can be found in \cite{Armatol:2021b,CROSSdeplLMO:2023}, for comparison). Taking into account a high performance of the $^{116}$CdWO$_4$ detector achieved in these measurements, we decided to keep it as a reference scintillating bolometer for the CROSS facility.

\subsubsection{Li$_2$MoO$_4$ scintillating bolometer}

For the construction of the Li$_2$MoO$_4$ scintillating bolometer we used the CLYMENE sample (Sec. \ref{sec:LMO_crystal}) and a LUMINEU-like detector holder made of Cu \cite{Armengaud:2017}, which was covered by layers of Au and Ag to increase light collection inside the cavity without using a reflective film, as seen in figure \ref{fig:Detectors} (Left, bottom). The Li$_2$MoO$_4$  bolometer was instrumented with two NTDs: 
one thermistor with a size of 3$\times$3$\times$1~mm was coupled with UV cured glue (Permabond), while the second sensor was about four time larger (4$\times$10$\times$1~mm) and glued using epoxy (Hysol 2039). In addition, a heating element was glued on the opposite side of the Li$_2$MoO$_4$ surface with the NTDs, as seen through the crystal in figure \ref{fig:Detectors} (Left, bottom). 
A circular Ge light detector, of the same size as one used in the $^{116}$CdWO$_4$ scintillating bolometer, was coupled to the Li$_2$MoO$_4$ crystal. The Ge wafer has Al concentric electrodes deposited on one surface, which can be used to amplify thermal signals exploiting the Neganov-Trofimov-Luke effect \cite{Novati:2019}. A small NTD (1$\times$2$\times$1~mm) has been epoxy-glued (with Araldite\textregistered Rapid) onto the Ge slab for the detection of thermal signals. 

It is worth mentioning that initially we constructed the Li$_2$MoO$_4$ bolometer being equipped with the large NTD only. At first this detector was tested above ground in a pulse-tube cryostat at IJCLab (Orsay), showing high performance, and then operated in the CROSS set-up (as seen in Fig.~2 of \cite{CROSSdeplLMO:2023}) with changed sensor properties (current–voltage characteristics with temperature) and degraded bolometric performance, indicating an issue with the sensor coupling to the crystal. Therefore, we refurbished the CLYMENE detector for the present study.

\section{Experiment in the CROSS underground facility}
\label{sec:Experiment}

Two modules of scintillating bolometers (described in Sec. \ref{sec:Detectors}) were installed inside the CROSS cryogenic facility, as seen in figure \ref{fig:Detectors} (Right). In this section we describe the CROSS set-up (Sec. \ref{sec:CROSS_facility}), its detector suspension upgrade (Sec. \ref{sec:Suspension}), and the measurements (Sec. \ref{sec:Measurements}).

\subsection{The CROSS facility}
\label{sec:CROSS_facility}

\begin{figure}
\centering
\includegraphics[width=1.0\textwidth]{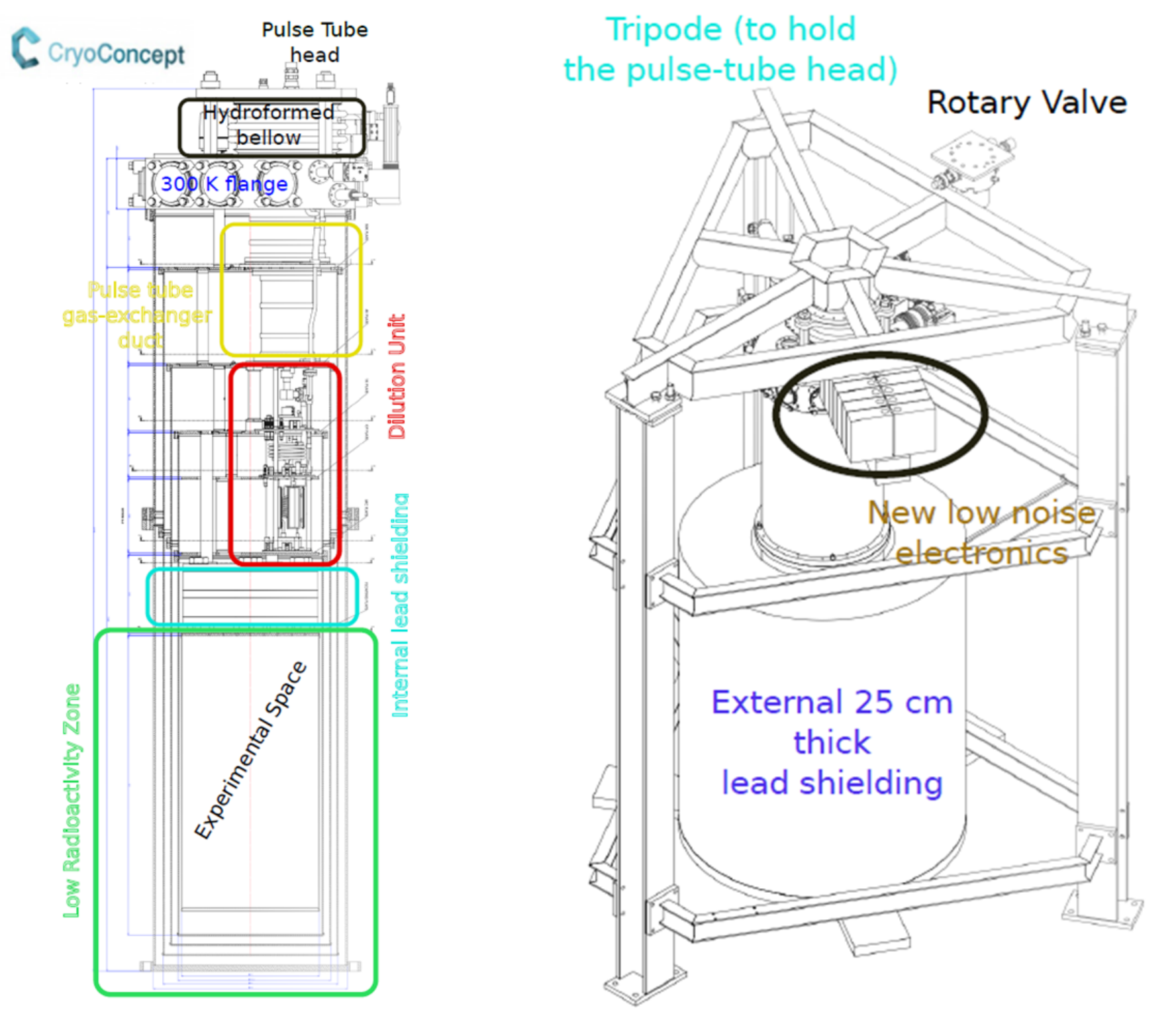}
\caption{Drawings of a pulse-tube dilution refrigerator (Left) developed in France by Cryoconcept and installed in the CROSS cryogenic facility (Right) hosted at LSC in Spain.}
\label{fig:CROSS_facility}
\end{figure}

The CROSS cryogenic facility (C2U) \cite{Olivieri:2020,Armatol:2021b} is hosted in the hall B of the Canfranc underground laboratory (LSC) in Spain \cite{Bettini:2012,Ianni:2016,Cebrian:2020a} and is actively used since its commissioning done in April--July 2019 \cite{Helis:2020,Helis:2021}. This facility provides a low-cost high-efficiency duty cycle thanks to the use of a dilution refrigerator (HEXA-DRY by CryoConcept) equipped with a pulse tube (Cryomech PT415) for cooling down to 4 K (see figure \ref{fig:CROSS_facility}, Left), thus avoiding a LHe-bath to be refilled periodically with expensive liquefied helium. Taking into account that such cryogenic instruments are known to be a notable source of vibrational noise \cite{Olivieri:2017}, the cryostat is assisted by the Ultra-Quiet Technology{\texttrademark} (UQT) \cite{UQT} to decouple the pulse tube from the dilution unit. To do so, the pulse tube head is coupled to the cryostat via a supple hydroformed bellow and is firmly kept by a tripode rigid frame (see figure \ref{fig:CROSS_facility}). While, the cryostat is mechanically tighten to a 10 ton lead-brick loaded platform. 
The pulse tube cold stages are thermally linked to the cryostat cold stages by gas loaded heat exchangers preventing any mechanical contacts. 
The dilution unit has a cooling power of about 340~$\mu$W at 100 mK and 5.5 $\mu$W at 20 mK (0.2 $\mu$W at 10 mK), and can reach a base temperature $\sim$10 mK over an experimental volume of 600 cm length and 30 cm diameter. The cryostat cools down to the base temperature over around 60 h; the regulated temperature of the cryostat fluctuates within 10 $\mu$K.

The CROSS facility is placed inside a Faraday cage with acoustic isolation, formerly used by the ROSEBUD dark matter experiment \cite{Cebrian:2004}. 
Thanks to the underground location of the set-up (the rock overburden of the LSC is around 2450 m water equivalent), a flux of cosmic-ray muons is significantly reduced compared to the surface level \cite{Trzaska:2019}. In order to suppress environmental $\gamma$ radiation, the outer vacuum chamber of the cryostat is surrounded by a 25 cm thick shielding made of a low-radioactivity lead (but still with a notable content of $^{210}$Pb \cite{CROSSdeplLMO:2023}). 
Additionally, the experimental volume of cryostat is protected internally from the top by a 13 cm thick disk made of interleaved lead and copper (about 120 kg), cooled down to temperatures as low as 800 mK. Furthermore, annular lead layers of about 8 mm thickness are firmly anchored to the 50 K, 4 K, and 1 K thermal screens to reduce a line-of-sight $\gamma$ flux from the top. Moreover, a stainless steel box is installed around the cryostat and flushed with a deradonized air flow ($\sim$1 mBq/m$^3$ of Rn \cite{PerezPerez:2022}), drastically reducing the radon induced background \cite{CROSSdeplLMO:2023}. Finally, the shielding of the CROSS set-up is completed with an outer muon veto made of plastic scintillators with SiPM readout, which is under commissioning now.

Seven over eight planned 12-channel boards of low-noise room-temperature DC front-end electronics \cite{Carniti:2020,Carniti:2023}, restyled from the Cuoricino experiment \cite{Arnaboldi:2002}, have been already plugged to the cryostat connector box. The new cards are composed with the digital control circuit, integrated 24-bit ADC and a programmable 6-pole Bessel-Thomson anti-aliasing filter. The data acquisition (DAQ) program is a MATLAB based graphical user interface, which acquires the continuous data; the data transfer to a DAQ computer is realized via 1 GB/s Ethernet (optically decoupled to completely insulate the DAQ computer). 
The cryostat wiring is done with twisted pair manganin wires; currently 84 over 168 planned channel lines are installed from 300 K connectors down to the mixing chamber, while only 36 channels are reaching the floating detector plate.  Also, a batch of optic fibres (64 lines in total) has been installed, from the room temperature feedthrough down to the experimental volume (52 fibres are fixed on 4 K plate and 12 are reaching the floating plate), in order to flush detectors with a burst of photons emitted by a room-temperature LED (880 nm of the photon emission maximum). The excitation of LED and heater (through a Si:P heater) signals is driven by a two-channel wave-function generator (Keysight 33500B). The data taking stability and quality is monitored on-line, while a routine data analysis is done off-line (as detailed below).

\subsection{New suspension with magnetic dumpers}
\label{sec:Suspension}

\begin{figure}
\centering
\includegraphics[width=0.8\textwidth]{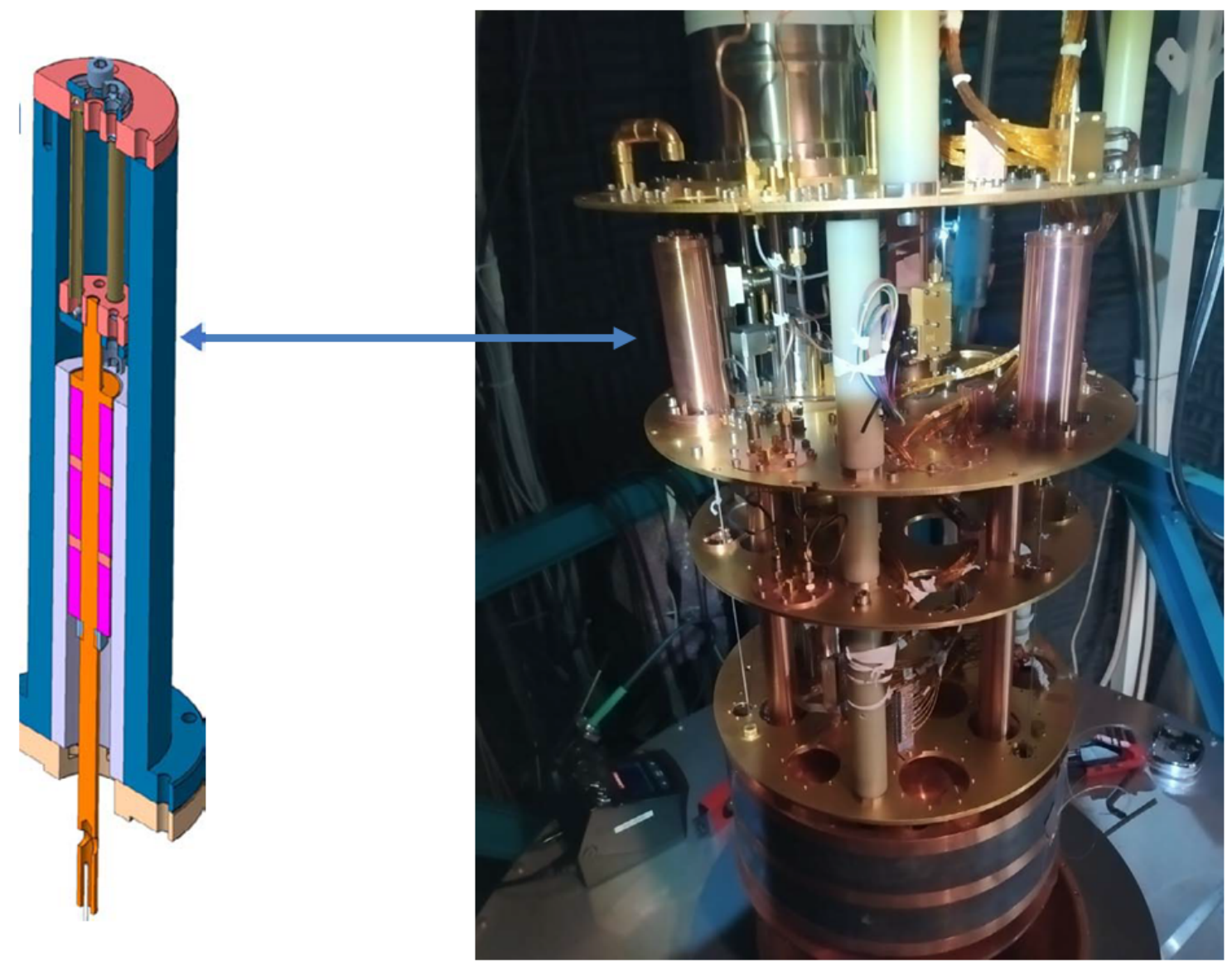}
\caption{Rendering of a magnetic dumper (Left), three of which are installed at the 1 K stage of the CROSS cryostat (Right). The detector plate is then suspended with three 40 cm long, 1 mm diameter Dyneema{\textregistered} ropes connected to the dampers via three springs.}
\label{fig:Magnetic_dumpers}
\end{figure}

The achievement of the optimal detector performance in a pulse-tube cryostat requires the development of an efficient mechanical decoupling system to prevent vibrations from spoiling the detector performance  \cite{Olivieri:2017}. All the decouplings realized so far in the CROSS set-up \cite{Armatol:2021b,CrossCupidTower:2023a,CROSSdeplLMO:2023,CupidAlternativeStructure:2023a} were provisional, consisting simply in stainless steel spring(s) and thermalization bands, all inside the experimental space at 10--20 mK. The major problems of these systems are that the cut-off frequencies are relatively high (a few Hz), especially the horizontal ``pendulum'' one, and that no controlled dumping is introduced.

Prior the measurements described in the present work, the final CROSS suspension system was installed. 
The new approach consists of suspending the full vessel that contains the detector (i.e. a top plate and a cylindrical Cu screen) at the base temperature. The top plate hangs from three Kevlar{\textregistered} ropes mechanically connected at the still plate (1 K), about 40 cm above the detector area.
Inspired from a recent published design \cite{Wang:2020}, the springs and Dyneema{\textregistered} ropes are anchored to ring-shaped NdFeB permanent magnets, which are left free to move vertically inside fixed and close Cu tubes (figure \ref{fig:Magnetic_dumpers}, Left). Their displacement provides an efficient dumping effect to the oscillations, thanks to the eddy currents created. The new detector suspension system of the CROSS facility is shown in figure \ref{fig:Magnetic_dumpers} (Right).

\subsection{Low-temperature operation of the detectors}
\label{sec:Measurements}

The full assembly of the $^{116}$CdWO$_4$ and Li$_2$MoO$_4$ scintillating bolometers was directly connected to the newly suspended detector plate of the CROSS cryostat (see figure \ref{fig:Detectors}, Right). The plate temperature was regulated at 15 mK. In order to find an optimal NTD polarization for each detector (providing the best signal-to-noise ratio), we investigated heater/LED induced response of bolometers varied with NTD current (e.g. see in \cite{Novati:2019}). 

After defining the optimal working points, we realized a set of measurements with and without a $^{232}$Th $\gamma$ source. The used source is made of a thoriated tungsten wire which was inserted inside the lead shield for calibration runs. A part of the data collected was dedicated to studies of pile-up rejection efficiency for next-generation $^{100}$Mo-based bolometric $0\nu2\beta$ experiments; an interested reader can find details and results of this investigation in \cite{CROSSpileup:2023}. Unfortunately, after about one month of the operation the cryogenic run has been terminated due to the failure of the primary circulation pump (already replaced). It does not have an impact on the research program planned, except of the lack of a light detector dedicated calibration with the high-intensity $^{232}$Th source.

The continuous data were digitized with a sampling rate of either 2 kS/s or 5 kS/s and the cut-off frequency of the low-pass Bessel filter was set between 300 Hz and 600 Hz for different datasets (here we present results for the 2 kS/s sampling and 300 Hz Bessel cut-off).  
The analysis of the collected data and the interpretation of the achieved results are presented in the next section (Sec. \ref{sec:Results}).

\section{Data analysis and results}
\label{sec:Results}

\subsection{Data processing}

We performed the processing of the acquired data using a MATLAB-based GUI tool developed at IJCLab \cite{Mancuso:2016}. In order to improve the signal-to-noise ratio, the data processing exploits the Gatti-Manfredi optimum filter \cite{Gatti:1986}. The use of the filter requires only two inputs characterizing the data. The first input is an average pulse-shape of a signal, which is created by superimposing normalized individual signals of tens high-energy events. The second input, is an average noise spectral density, which represents noise fluctuations in the frequency domain observed in a large number (10000) of waveforms containing no signal. The optimum filter is used for both triggering of events and the final processing. The energy threshold of 5 to 10 RMS of the baseline noise fluctuation is typically set to trigger events. The output of the program contains an event-by-event information about the signal amplitude (i.e. energy) and calculated several pulse-shape parameters. 

Two pulse-shape parameters relevant for the present work describe time constants of a bolometric signal: 1) rise time defined as the time required by the signal to increase from 10\% to 90\% of its height; 2) decay time, the time required to decrease from 90\% to 30\% of its height. Another two parameters typically used for events selection are a correlation (a Pearson’s linear correlation coefficient between signals of a triggered event and a template) and a normalized amplitude (a ratio between the signal amplitudes extracted after an optimum filtering and fit) described in detail in  \cite{Bandac:2020}. 

In order to establish coincidences between the scintillating bolometers and accompanied light detectors, we used triggers of the former channels to process the latter ones and accounting for a difference in the time response, similarly to the method described in \cite{Piperno:2011}. It is worth noting that the crystals scintillation was used for calibration of light detectors (see e.g. in \cite{Novati:2019}), taking into account the knowledge about the light signal relative to heat energy release, measured for the used detector modules with dedicated light detector calibrations in the previous tests.

\subsection{Detectors performance}

The results on the $^{116}$CdWO$_4$ and Li$_2$MoO$_4$ scintillating bolometers characterization are summarized in table \ref{tab:Performance} together with the performance of the $^{116}$CdWO$_4$ reference detector reported for its previous operation in the CROSS set-up before the new suspension implementation.


\begin{table}
 \caption{Performance of $^{116}$CdWO$_4$ and Li$_2$MoO$_4$ scintillating bolometers; the results of the former module tested in the CROSS set-up before the suspension upgrade \cite{Helis:2020,Helis:2021} are given for comparison. We report the NTD size and its resistances ($R_{NTD}$) at a given current ($I_{NTD}$), the rise and decay time parameters ($\tau_R$ and $\tau_D$, respectively), the detector sensitivity ($A_{Signal}$), and the energy resolution of the baseline noise and the 2615 keV $\gamma$-ray peak of $^{208}$Tl from the $^{232}$Th source (an improvement by accounting heat-light anti-correlation is labeled with $^a$). The LD-LMO performance enhanced with the Neganov-Trofimov-Luke signal amplification (60 V electrode bias) is marked with $^b$.}
\footnotesize
\begin{center}
\begin{tabular}{l|c|c|c|c|c|c|c|c|c}
 \hline
Detector & Suspension & NTD & $R_{NTD}$ & $I_{NTD}$ & $\tau_R$ & $\tau_D$ & $A_{Signal}$ & \multicolumn{2}{c}{FWHM (keV)}  \\
~ & ~ & size (mm) & (M$\Omega$) & (nA) & (ms) & (ms) & ($\mu$V/keV) & noise & 2615 keV $\gamma$  \\
\hline
\hline
CWO      & No  \cite{Helis:2020,Helis:2021} & 3$\times$3$\times$1 & - & 2.5   & -   & -   &  0.036       & 10.1  & 16.1(3) \\
~        & 1-spring \cite{Helis:2021} & ~ & 10 & 1.0   &  -  &  -  &  0.081   & 3.1   & 8.8(3) \\
~        & New                 & ~ & - & 2.0   & 16 & 286  &  0.039  & 1.7   & 7.0(3) \\
~        & ~                   & ~ & ~ & ~  & ~   & ~   & ~        & ~      & 6.2(2)$^a$ \\
\hline
LMO      & New                 & 4$\times$10$\times$1 & 6.4 & 3.3    & 6.2   & 41   & 0.032    & 8.9        & -   \\
~        & ~                   & 3$\times$3$\times$1 & 9.8 & 2.0    & 15  & 67   & 0.029    & 3.1   & -       \\
~        & ~                   & ~ & - & 0.28   & 23  & 80  & 0.14    & 2.3   & 6.0(5) \\

\hline
\hline
LD-CWO    & No  \cite{Helis:2020,Helis:2021}  & 1$\times$3$\times$1 & -   & 1.3   & -   & -   & 1.1    & 0.36 & n.a. \\
~ & 1-spring \cite{Helis:2021}  & ~ & 3.0 & 2.9  & -  & -   & 1.9  & 0.22  & n.a.  \\ %
~ & New   & ~ & 2.3 &  1.0 & 2.0  & 4.3  & 2.7  & 0.10  & n.a.  \\ %
\hline
LD-LMO    & New  & 1$\times$2$\times$1 & 2.1 & 1.0  & 1.3  & 4.0   & 1.6   & 0.13  & n.a.  \\ 
~         & ~  & ~ & ~ & ~  & 1.2  & 4.5   & 11$^b$   & 0.018$^b$  & n.a.  \\ 
~         & ~  & ~ & - & 0.28  & 2.1  & 4.2   & 34$^b$   & 0.016$^b$  & n.a.  \\ 
\hline
 \end{tabular}
  \label{tab:Performance}
 \end{center}
 \end{table}

\normalsize

\nopagebreak

\subsubsection{$^{116}$CdWO$_4$ bolometer}

The NTD of $^{116}$CdWO$_4$ thermal detector for the chosen working point was polarized with a relatively high current (2 nA), thus reducing the detector sensitivity to around 40 nV/keV (see in table \ref{tab:Performance}), to the value which is not exceptional at 15 mK. 
The signal rise time is relatively fast, 16 ms, while the descending part of the signal is comparatively long, as exhibited by almost 300 ms decay time. The baseline energy resolution of our reference detector achieved in the upgraded CROSS set-up is a factor 2 improved, for the twice lower sensitivity, compared to the best previous measurement realized in this facility with a single-spring-suspended module \cite{Helis:2021}.

\begin{figure}[hbt]
\centering
\includegraphics[width=0.49\textwidth]{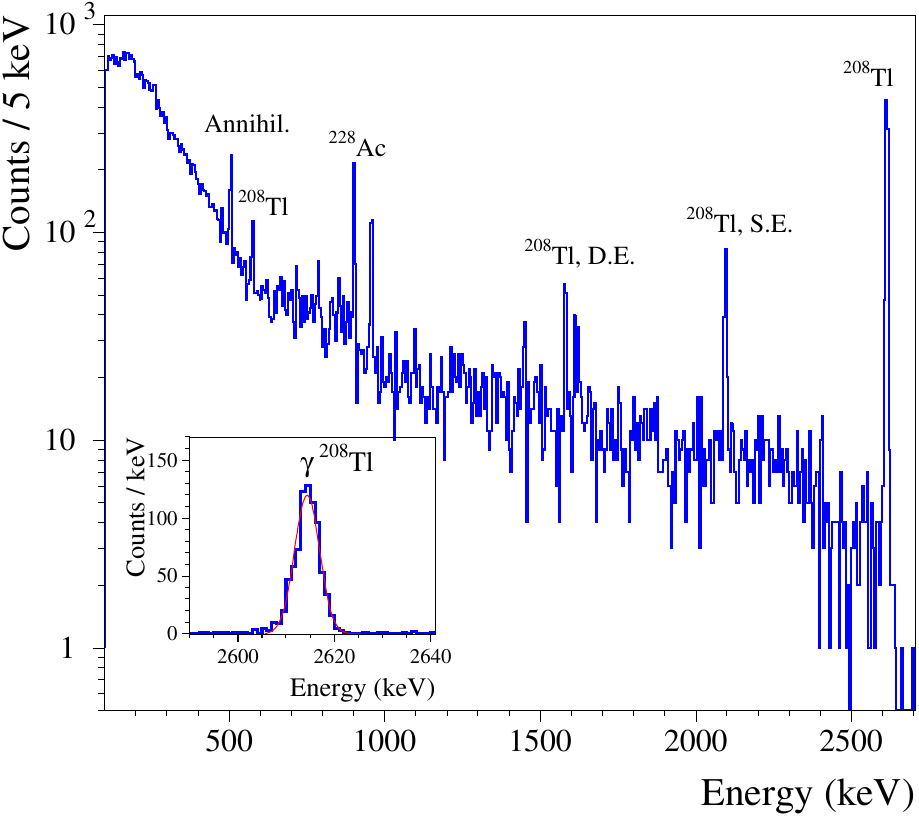}
\includegraphics[width=0.49\textwidth]{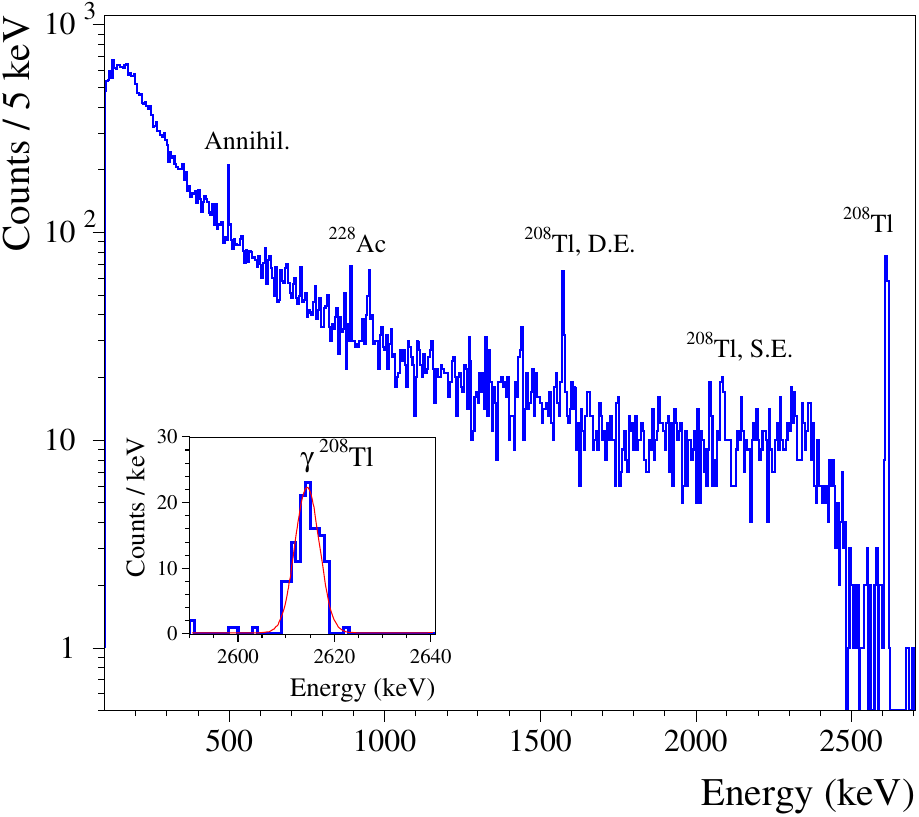}
\caption{Energy spectra of a $^{232}$Th source measured in a 72 h long calibration run of $^{116}$CdWO$_4$ (Left) and Li$_2$MoO$_4$ (Right) bolometers, based on large-volume (about 70 cm$^3$) crystal scintillators. The origin of the most intense $\gamma$ peaks of the two spectra is labeled (D.E. and S.E. are double and single escape peaks, respectively). The difference in the rate of $\gamma$ source induced peaks, evident from the comparison of the spectra, clearly shows the impact of the material density and the effective atomic number on the $\gamma$ detection efficiency. The inset of each figure shows a 2615 keV $\gamma$ peak of $^{208}$Tl together with a Gaussian fit (red line) returning the energy resolution of 6.2(2) and 6.0(5) keV FWHM for $^{116}$CdWO$_4$ and Li$_2$MoO$_4$ bolometers, respectively.}
  \label{fig:Th_spectra}
\end{figure}

Thank to a high density and high $Z_{eff}$ (an effective atomic number) of the cadmium tungstate, the detector exhibits intense $\gamma$ peaks in the relatively short datasets (tens of hours), as illustrated in figure \ref{fig:Th_spectra} (Left). The most energetic $\gamma$ peak, induced by 2615 keV $\gamma$ quanta from $^{208}$Tl decay, was measured with the energy resolution of 7.0(2) keV FWHM, which is almost 2 keV better than the resolution of the 1-spring-suspended module, and a factor of 2 better than the results of the measurements done without a dedicated suspension (see in table \ref{tab:Performance}). Moreover, knowing that the energy resolution of a cadmium tungstate crystal operated as thermal detector can be affected by anti-correlation between the heat and scintillation energy release \cite{Arnaboldi:2010a}, we corrected this effect, thus slightly improving energy resolution to 6.2(2) keV FWHM. This result is the same to the best ever reported energy resolution of a massive CdWO$_4$ thermal detector \cite{Arnaboldi:2010a,Poda:2017a}. 
We would like to emphasize that, in agreement with our early study with the sample produced from the same $^{116}$CdWO$_4$ ingot \cite{Barabash:2016} and in contrast to early studies of CdWO$_4$ bolometers \cite{Gironi:2009,Arnaboldi:2010a}, we see a minor impact of the heat-light anti-correlation on the detector energy resolution, which can be explained by the higher crystal quality of the samples used in our studies, which was achieved thanks to a dedicated purification of the starting materials and an advanced crystallization \cite{Barabash:2011}.

\subsubsection{Li$_2$MoO$_4$ bolometer}

The study of the static properties of NTDs, done during the working point optimization, shows that the resistance of the large NTD of the LMO detector being polarized at low current (hundreds pA) is about 100 M$\Omega$, which is a high level for our low-impedance electronics. Therefore, we polarized both NTDs of this detector at a few nA current to compare their performance, reported in table~\ref{tab:Performance}. We observed that the response of the large NTD sensor is about twice faster than that of the small one; it is explained by the difference in the glued area of the thermistors. At the same time, the sensitivity of both channels is almost the same, while the baseline noise is a factor 3 higher for the large NTD. 

In the test with NTD polarized at a low current (i.e. colder crystal working temperature) we grounded the large NTD channel due to a high resistivity of the sensor mentioned above. The other NTD channel demonstrated a factor of 5 gained sensitivity, while the baseline noise was improved by 25\% (see in  table~\ref{tab:Performance}). A ``modest'' improvement of the baseline noise compared with the largely gained sensitivity is explained by a Johnson noise contribution which becomes dominant for a high impedance NTD. The baseline noise resolution at this configuration, around 2 keV FWHM, is a good result for LMO-based thermal detectors, in particular for those operated in a pulse-tube cryostat. 

We also used a high sensitivity working point in a long calibration measurement aiming at precise investigation of the detector energy resolution; keeping in mind a low density and low $Z_{eff}$ of the material, which strongly affect the detection efficiency for $\gamma$ quanta, especially at higher energies which are of particular interest for this study. An example of the energy spectrum measured in one calibration dataset is shown in figure \ref{fig:Th_spectra} (Right). The energy resolution of the 2615 keV $\gamma$ peak, estimated from the Gaussian fit to each dataset, was found to be between 5.4(8) and 6.5(9) keV FWHM. The combined data corresponding to 271 h of measurements exhibits the energy resolution 6.0(5) keV FWHM of the 2615 keV peak. These results are among the best reported for Li$_2$MoO$_4$-based thermal detectors, in particular for those tested in the CROSS set-up.

\subsubsection{Bolometric light detectors}

\begin{figure}[hbt]
\centering
\includegraphics[width=1.0\textwidth]{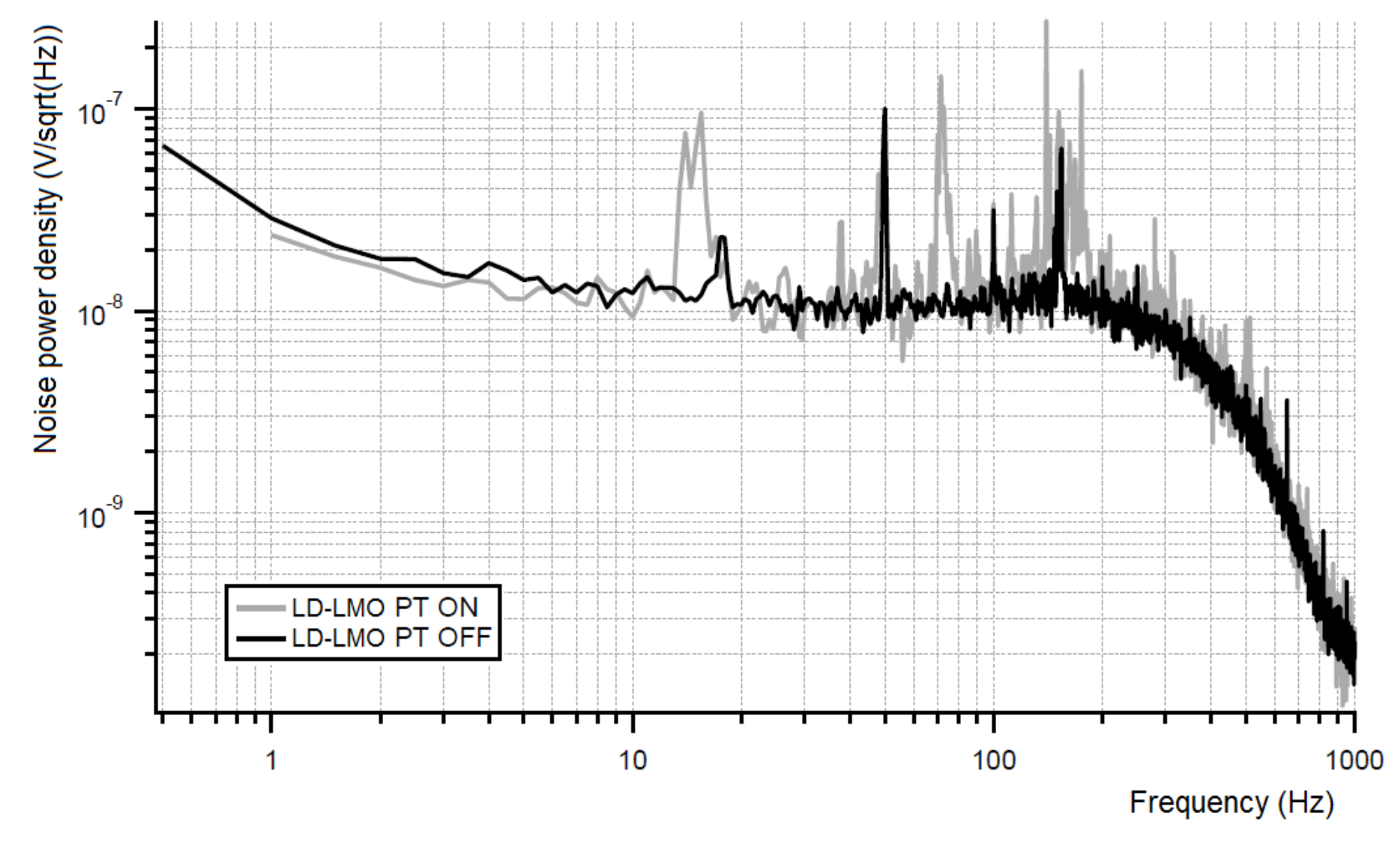}
\caption{Noise power spectral density as a function of frequency measured with a bolometric Ge light detector (LD-LMO) at 20 mK detector plate temperature and with a pulse tube switched ON (grey) and OFF (black). For both measurements, a 60 V electrode bias was set to operate the detector in the Neganov-Trofimov-Luke amplification mode.}
  \label{fig:NPS_Geco2}
\end{figure}

Ge light detectors being large-area low-mass bolometric devices are rather sensitive to mechanical vibrations induced by a pulse-tube cryostat (e.g. see in \cite{Armengaud:2017,Novati:2019}). In the present work we studied two detectors with the same 3 PTFE-based support of the Ge wafer, with the difference in the presence or not an individual Cu casing (see in figure \ref{fig:Detectors}, Left). Before cooling down to 15 mK, we did preliminary measurement at 20 mK and we observed that despite an upgrade with a new suspension, the pulse-tube induced vibration peaks (harmonics of the main 1.4 Hz frequency) remain in the signal bandwidth above 10 Hz, as illustrated in figure \ref{fig:NPS_Geco2}. Similar noise patterns were also observed in the previous measurements \cite{CrossCupidTower:2023a}, before the upgrade.

Calibration of the light detectors was realized using scintillation light emitted by the coupled crystal scintillators (see Sec. \ref{sec:PID}); the energy scale of the scintillation was measured in previous tests of these detector modules with LDs being calibrated with X-rays ($^{116}$CdWO$_4$ at LSC \cite{Helis:2020} and Li$_2$MoO$_4$ at IJCLab). It should be noted that we used another LD for Li$_2$MoO$_4$, germanium on sapphire, which was not SiO-coated and thus we assume at least 20\% improved light collection for the presently used device\footnote{Taking into account that SiO coating improves light collection by 20\%--30\% \cite{Mancuso:2014}, the uncertainty of the LD-LMO calibration is around 10\%.}.

The light detector of the reference module (LD-CWO) was characterized by a high sensitivity of 3 $\mu$V/keV; a 1--2 $\mu$V/keV signal was measured in the previous tests of this device. The achieved noise resolution, $\sim$100 eV FWHM, is a factor 2--3 better than early reported values \cite{Helis:2021}. However, similar and even better noise resolution (60--90 eV FWHM) has been achieved with similar size and performance (2.0--2.7 $\mu$V/keV sensitivity) bolometric light detectors tested in the CROSS set-up with a provisional suspension \cite{CrossCupidTower:2023a}. This suggests that the pulse-tube induced noise seen in the data (as exampled in figure \ref{fig:NPS_Geco2}) is picked-up by wires in uncontrolled way. 

In case of the light detector (LD-LMO) coupled to the Li$_2$MoO$_4$ bolometer, we enhanced its performance by applying a 60 V bias on the Al electrode; it allowed us to get a high sensitivity of about 10--30 $\mu$V/keV and a baseline noise of only 16-18 eV FWHM (depending on the working point). These results additionally demonstrate how powerful and viable is the Neganov-Trofimov-Luke signal amplification for the improvement of the signal-to-noise ratio of light detectors, which is of special interest for pile-up background rejection in bolometric $0\nu\beta\beta$ experiments \cite{Chernyak:2012,Chernyak:2014,Chernyak:2017,Armatol:2021,CROSSpileup:2023}.

\subsection{Particle identification of scintillating bolometers}
\label{sec:PID}

Cadmium tungstate is known to be an efficient scintillator at low temperatures, while lithium molybdate is characterised by a low scintillation efficiency, however still sufficient for particle identification (see in \cite{Poda:2020} and references therein). 

Indeed, in the early test of the $^{116}$CdWO$_4$ scintillating bolometer, with the light detector calibrated by an external X-ray $^{55}$Fe source, a relatively high energy of the scintillation signal with a mean of 25.2(1) keV was measured by the light detector per 1 MeV particle energy deposition in the $^{116}$CdWO$_4$ bolometer (for events with energies 0.5--2.7 MeV) \cite{Helis:2020}. This result is slightly lower than the record value of 31 keV/MeV achieved with the smaller $^{116}$CdWO$_4$ bolometer \cite{Barabash:2016}. Taking into account the high scintillation efficiency and a low noise of the light detector (see in table \ref{tab:Performance}), the $^{116}$CdWO$_4$ scintillating bolometer demonstrates a highly efficient particle identification capability, similar to the early reported results of this detector \cite{Helis:2020}.

\begin{figure}[hbt]
\centering
\includegraphics[width=0.75\textwidth]{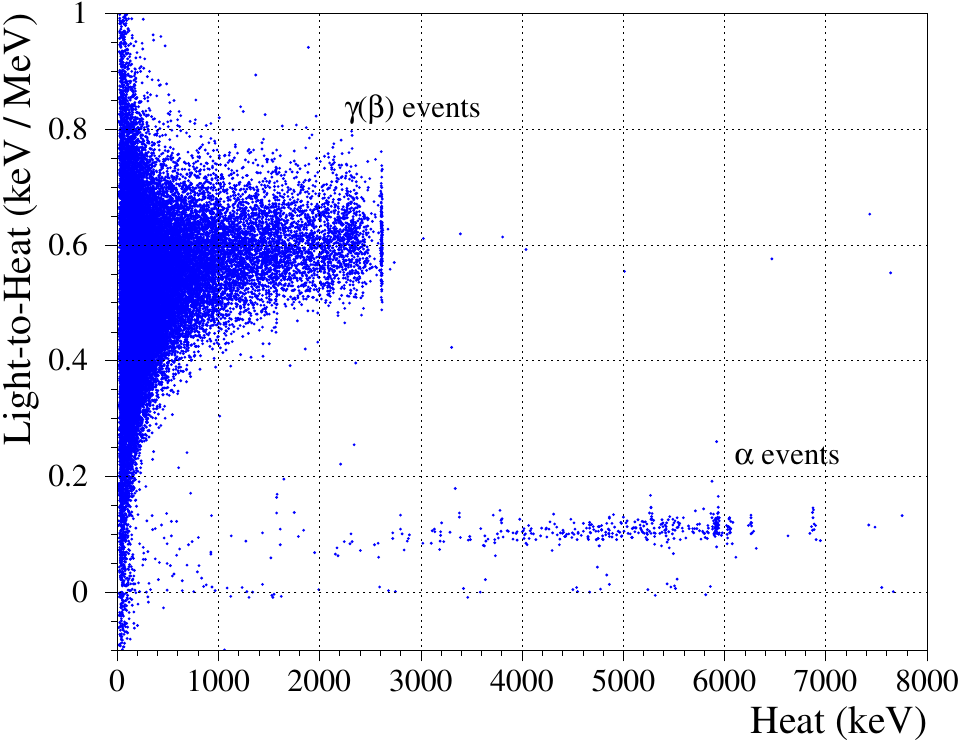}
\caption{Energy dependence of the light detector signals normalized on the corresponding heat energy release in the Li$_2$MoO$_4$ bolometer (light-to-heat parameter), measured over 271 h of calibrations with a $\gamma$ source of $^{232}$Th. Two populations of $\gamma$($\beta$) and $\alpha$ events are clearly separated thanks to the enhanced performance of the light detector, operated in the Neganov-Trofimov-Luke mode (see text).}
\label{fig:LMO_Particle_ID}
\end{figure} 


In contrast to the $^{116}$CdWO$_4$ detector, the Li$_2$MoO$_4$ scintillating bolometer in the aboveground test showed a low scintillation, which was evaluated to be around 0.6 keV/MeV scintillation signal (associated to 2615 keV $\gamma$ quanta of $^{208}$Tl). Such scintillation signal is typical for Li$_2$MoO$_4$ bolometers, e.g. 0.5--0.9 keV/MeV were reported depending on the light collection efficiency for detectors surrounded by a reflector (see in \cite{Poda:2020}); the absence of the reflector would reduce the measured scintillation signal by at least a factor 2 \cite{Armatol:2021a,CrossCupidTower:2023a}. If the baseline noise of a light detector is 100 eV FWHM and below, one can get a powerful particle identification despite a low scintillation efficiency, as e.g. demonstrated in \cite{CrossCupidTower:2023a}. In order to secure a low noise of light detectors and to improve their signal-to-noise ratio, the signal amplification based on the Neganov-Trofimov-Luke effect represents a viable upgrade of ``ordinary'' bolometric Ge light detectors with no impact on detector design and sensor technology, but with a price of additional wiring required to polarize electrodes deposited on light detectors, as e.g. in \cite{Novati:2019}. In particular, by applying 60 V on the electrode of the LD-LMO, we achieved a factor of 10 gain in the signal-to-noise ratio and the baseline energy resolution of 16 eV FWHM. Thanks to the drastic enhancement of the light detector performance, we get a highly efficient particle identification, compatible to the results of $^{116}$CdWO$_4$ detectors \cite{Barabash:2016,Helis:2020}, as illustrated in figure \ref{fig:LMO_Particle_ID}. A high sensitivity of the light detector also allowed to demonstrate a minor impact of non-linearity of scintillation output from the Li$_2$MoO$_4$ crystal on the collected data, exhibited in figure \ref{fig:LMO_Particle_ID} as a slope of a mean value of the heat-normalized light signal for high energy events detected by the Li$_2$MoO$_4$ bolometer.

We would like to point out on the detection of some $\alpha$ events with almost zero light associated, despite using the high-sensitivity bolometric photodetector. Moreover, the thermal energy of these ``dark'' $\alpha$ events is higher than that of $\alpha$s with well-detectable light signals. It indicates on the impact of small defects (bubbles) in the crystal on the short-range interaction particles, like $\alpha$s. Such behaviour is similar to zinc molybdate bolometers based on crystals with visible structure defects \cite{Armengaud:2017}. It is worth emphasizing that, having no impact on long-range interaction particles like electrons, the observed effect does not affect a double-beta decay signal. Also, it should be noted that only a single crystallization was performed for the CLYMENE crystal, while the re-crystallization was adopted for the LUMINEU technology to improve both radiopurity and quality of Li$_2$MoO$_4$ crystals \cite{Armengaud:2017,Grigorieva:2017}.

\subsection{Radiopurity of the CLYMENE crystal}

Data acquired without the $^{232}$Th source were used to investigate radioactive contamination of only Li$_2$MoO$_4$ crystal ---a viable input for the CLYMENE technology of the crystal production--- while, the radiopurity of the $^{116}$CdWO$_4$ crystal has been already extensively studied by operating it as a room-temperature scintillation detector \cite{Barabash:2011,Barabash:2018} and as a cryogenic bolometer \cite{Helis:2020}.

\subsubsection{Analysis of $\alpha$ events}

\begin{figure}[hbt]
\centering
\includegraphics[width=0.75\textwidth]{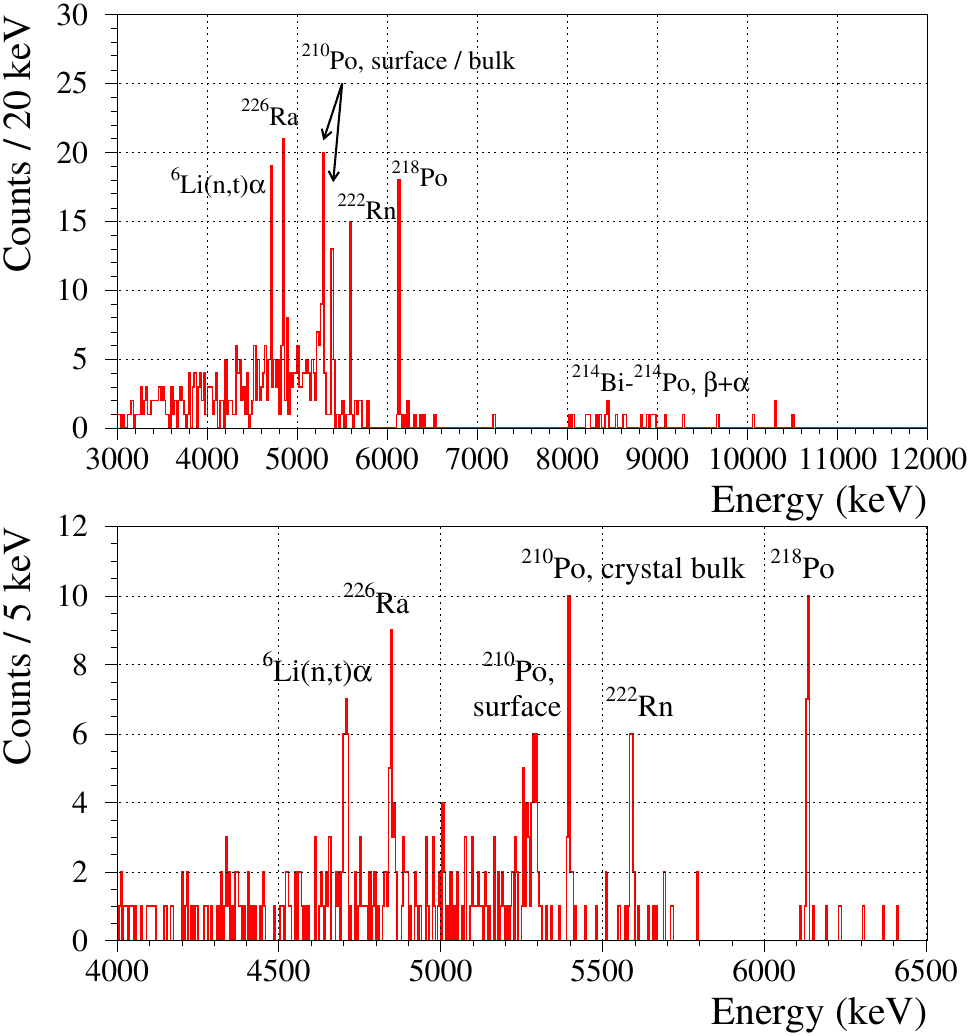}
\caption{Fragments of the energy spectrum of $\alpha$ events acquired by a thermal detector based on a 245 g Li$_2$MoO$_4$ crystal scintillator in a 244 h long measurement in the CROSS underground facility. We show 3.0--12 MeV (top) and 4.0--6.5 MeV (bottom) intervals covering the expected energy of $\alpha$-active radionuclides from U/Th decay chains.}
\label{fig:LMO_Bkg_alpha}
\end{figure}

Thanks to a complete separation between $\gamma$($\beta$) and $\alpha$ events achieved with the Li$_2$MoO$_4$ scintillating bolometer, as illustrated in figure \ref{fig:LMO_Particle_ID}, we selected the distribution of $\alpha$s to analyze a possible $\alpha$-activity of radionuclides from U and Th decay families. The selection efficiency (includes the trigger efficiency) has been investigated using a pulse-injection method as described e.g. in \cite{CROSSpileup:2023}; we found the efficiency of 97\% in the region of interest for $\alpha$ particles. The energy spectrum of the selected $\alpha$ events is shown in figure \ref{fig:LMO_Bkg_alpha}.

The data of the CLYMENE Li$_2$MoO$_4$ crystal clearly exhibit several peaks. The lowest energy peak seen in figure \ref{fig:LMO_Bkg_alpha} is ascribed to the products of neutron capture by $^6$Li, a summed energy of alpha particle and triton emitted in the reaction ($Q$-value is 4784~keV). The counting rate of such events is 1.9(4)~counts/d, in a good agreement with our expectation ($\sim$1.8~counts/d) based on the results of the similar-size Li$_2$MoO$_4$ bolometers operated in the C2U set-up \cite{CROSSdeplLMO:2023,CrossCupidTower:2023a}. 

The next to the neutron capture signature is a $^{226}$Ra peak ($Q_{\alpha}$ = 4871~keV), which is accompanied by the two peaks originated to $\alpha$ decays of daughter radionuclides, $^{222}$Rn ($Q_{\alpha}$ = 5590~keV) and $^{218}$Po ($Q_{\alpha}$ = 6115~keV), and a continuum of events in the energy interval of 7.8--11~MeV corresponding to the summed energy deposition induced by $\beta$ and subsequent $\alpha$ decays of $^{214}$Bi ($Q_{\beta}$ = 3269~keV) and $^{214}$Po ($Q_{\alpha}$ = 7833~keV, $T_{1/2}$ = 164~$\mu$s), respectively. Moreover, we found that most ($\sim$85\%) of $^{222}$Rn candidates selected in the energy interval 5520--5740 keV were followed in a short time by $^{218}$Po candidates detected with 6100--6450~keV energy. The time difference between two subsequent $\alpha$ events is between 10 s and 1000 s, i.e. up to 5$\times$$T_{1/2}$ of $^{218}$Po ($T_{1/2}$ = 3.10~min). Thus, we clearly see the internal contamination of the  CLYMENE Li$_2$MoO$_4$ crystal by $^{226}$Ra with the activity of 0.12(2)~mBq/kg.

Also, we see a $^{210}$Po doublet: the sharp peak corresponds to the signature of $^{210}$Po decay ($Q_{\alpha}$ = 5407~keV) in the crystal bulk, while the broad peak with the mean energy of 5.3~MeV is originated to decays of $^{210}$Po on the detector surfaces thus only $\alpha$ particles (without nuclear recoils of $^{206}$Pb) were detected. 
The 5.3 MeV $^{210}$Po peak is accompanied by a tail of events with lower energies, which suggests a surface contamination of the crystal probably by the polishing products used (alumina, silicon carbide). The internal $^{210}$Po activity is measured as 0.09(2)~mBq/kg.

\subsubsection{Study of $\gamma$($\beta$) events}

\begin{figure}[hbt]
\centering
\includegraphics[width=0.75\textwidth]{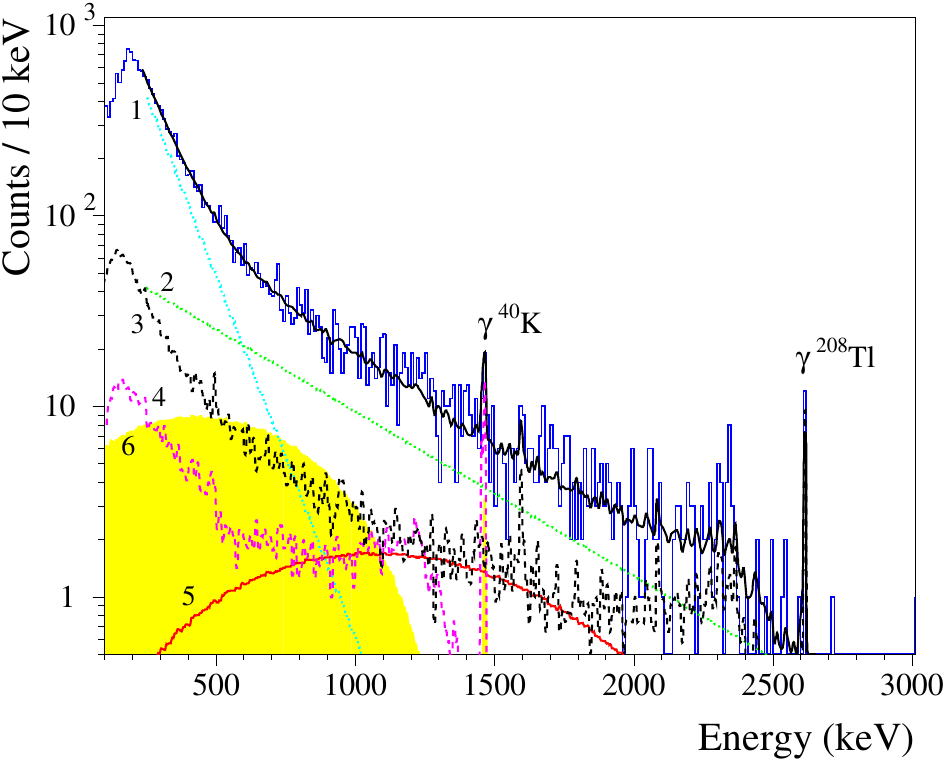}
\caption{Energy spectrum of $\gamma$($\beta$) events detected by the 245~g Li$_2$MoO$_4$ scintillating bolometer over 244~h of measurements in the CROSS facility at the Canfranc underground laboratory. The prominent $\gamma$ peaks of $^{40}$K and $^{208}$Tl are labeled. Fit by a simplified background model is shown by a solid black line and the background components (see text) are labeled as follows: 1 -- $^{210}$Bi induced Bremsstrahlung $\gamma$ rays; 2 -- other external $\gamma$ quanta except $^{232}$Th and $^{40}$K; 3 -- external $\gamma$ from $^{232}$Th; 4 -- external $\gamma$ from $^{40}$K; 5 -- $2\nu\beta\beta$ decay of $^{100}$Mo in the crystal; 6 -- internal $^{40}$K.}
\label{fig:LMO_Bkg_gamma}
\end{figure}

In addition to the study of the $\alpha$ spectrum, we investigated $\gamma$($\beta$) events measured by the Li$_2$MoO$_4$ scintillating bolometer aiming at probing a possible content of $^{40}$K, a well known issue of the lithium molybdate material taking into account the chemical affinity of K and Li (e.g. see in \cite{Barinova:2009,Armengaud:2017}). However, the dominant continuum distribution of internal $^{40}$K complicates the study compared to the $\alpha$ spectroscopy and it requires a dedicated background model. Thus, keeping in mind results of the previous measurements with Li$_2$MoO$_4$-based bolometers in the C2U set-up \cite{CROSSdeplLMO:2023}, we constructed a simplified background model to take into account the following contributions: 
\begin{itemize}
  \item  Bremsstrahlung $\gamma$ rays from $^{210}$Bi decays (originated to the $^{210}$Pb content in the lead shield) modeled by an exponential function;
  \item $\gamma$ quanta from $^{232}$Th and $^{40}$K decays in the external sources, modeled as measured energy spectra with such calibration sources. The $^{232}$Th model was build using the CLYMENE detector data of calibrations with a low-intensity source (130 h), while the $^{40}$K calibration data of similar size and performance two Li$_2$MoO$_4$ thermal detectors (20 h each) \cite{Armengaud:2017} were used to construct the model of external $^{40}$K;
  \item decays of $2\nu\beta\beta$ decay of $^{100}$Mo (with the expected activity of 1 mBq/kg \cite{Armengaud:2020b}) and $^{40}$K (effect searched for) in the Li$_2$MoO$_4$ crystal, simulated using the GEANT4-based code \cite{Kobychev:2011} and the DECAY0 event generator \cite{Ponkratenko:2000};
  \item residual background, not accounted by the above listed components, also modeled by an exponential function.
\end{itemize}
\noindent Using the simplified model, we performed a least-squares fitting to the data in the energy interval 230--2650 keV, achieving a good description of the measured features of the spectrum ($\chi^2$/n.d.f. = 212/209 = 1.01), as shown in figure \ref{fig:LMO_Bkg_gamma}. We see an indication on the bulk $^{40}$K activity; the fit returns ($783 \pm 266$) decays of $^{40}$K corresponding to ($3.8 \pm 1.3$)~mBq/kg activity. Taking into account the evident imperfection of our simplified background model, as well as a large uncertainty of the observed $^{40}$K signal, we prefer to give an upper limit on the $^{40}$K activity in the crystal. Thus, using the Feldman-Cousins approach~\cite{Feldman:1998}, we get an upper limit (1219 counts at 90\% C.L.) on decays which can be ascribed to the $^{40}$K radioactivity in the crystal bulk and set an upper limit on its activity as $\leq$6~mBq/kg at 90\% C.L.

\subsubsection{Comparison of the crystal radiopurity}

Summarizing the radiopurity study with the CLYMENE crystal, presented in table\ref{tab:Radiopurity}, we can conclude that the activity of $^{210}$Po in the sample is on the similar level of this contamination in the Li$_2$MoO$_4$ crystals produced for LUMINEU \cite{Armengaud:2017,Poda:2017a}, CUPID-Mo \cite{Armengaud:2020a,Poda:2020} and CROSS \cite{Armatol:2021b,CrossCupidTower:2023a} experiments, and is acceptable for next-generation experiments like CUPID. The $^{210}$Po content is originated to a $^{210}$Pb contamination typical for crystal scintillators and particularly for molybdates \cite{Armengaud:2015,Armengaud:2017,Danevich:2017,Danevich:2018}. 
The $^{226}$Ra activity in the CLYMENE crystal is about an order of magnitude higher than the objectives for next-generation experiments, while we do not see a signature of $^{228}$Th and its activity should comply with such demands. Also, the presence of $^{40}$K on a dangerous level (tens mBq/kg, becoming an additional dominant source of pile-ups \cite{Armengaud:2017}) is excluded by the CLYMENE crystal data and the upper limit on the $^{40}$K activity is similar to the results of LUMINEU crystals ($\leq$3 mBq/kg) with comparable exposure \cite{Armengaud:2017}.

\begin{table}
 \caption{Radioactive contamination of the Li$_2$MoO$_4$ crystal (produced within the CLYMENE project). The uncertainties are given at a 68\% C.L., while the limits are set at 90\% C.L. Radiopurity of the material, produced for LUMINEU, CUPID-Mo and CROSS $\beta\beta$ experiments and studied over a similar exposure, is given for comparison.}
\footnotesize
\begin{center}
\begin{tabular}{c|c|c|c|c|c}
 \hline
Chain & Nuclide & \multicolumn{4}{c}{Activity (mBq/kg) in Li$_2$MoO$_4$ with different $^{100}$Mo content}  \\
\cline{3-6}
~ & ~ & \multicolumn{2}{c|}{Natural composition} & Enriched in $^{100}$Mo & Depleted in $^{100}$Mo \\
\cline{3-4}
~ & ~ & Present work & \cite{Armengaud:2017} & \cite{Armengaud:2017,Poda:2017a,Armengaud:2020a,Poda:2020,Armatol:2021b,CrossCupidTower:2023a} & \cite{CROSSdeplLMO:2023} \\
\hline
$^{232}$Th & $^{228}$Th & $\leq$0.03 & $\leq$0.02 & $\leq$(0.003--0.006) & $\leq$0.002 \\
\hline
$^{238}$U & $^{226}$Ra & 0.12(2) & $\leq$0.04 & $\leq$(0.001--0.01) & $\leq$(0.004--0.007) \\
~ & $^{210}$Po & 0.09(2) & 0.1--0.2 & 0.01--0.4 & 0.04 \\
\hline
~ & $^{40}$K & $\leq$6 & $\leq$(3.2--12) & $\leq$3.5 & -- \\
\hline
 \end{tabular}
  \label{tab:Radiopurity}
 \end{center}
 \end{table}

\normalsize 

\nopagebreak

It is worth noting that initial, i.e. not additionally purified, starting materials were used for the Li$_2$MoO$_4$ crystal production \cite{Ahmine:2022a}, and this test is not intended to prove an ultra high radiopurity. At the same time, the bulk U/Th and K contamination of the tested crystal is rather low for a sample produced without a particular care on its radiopurity. These results give an idea of the radiopurity levels which can be reached with the purest, but unpurified, commercial powders used for the Li$_2$MoO$_4$ charge synthesis and crystal growth \cite{Ahmine:2022a}.

\section{Conclusions}

We investigated two scintillating bolometers, based on large-volume $^{116}$CdWO$_4$ and Li$_2$MoO$_4$ crystals, in the CROSS cryogenic facility at the Canfranc underground laboratory in Spain. The set-up has been upgraded with a new detector suspension system which implements a spring-suspended pendulum with magnetic dumpers installed at the 1 K stage of a pulse-tube dilution refrigerator. For the detector construction we used a $^{116}$CdWO$_4$ crystal scintillator, already operated in the CROSS set-up, and a Li$_2$MoO$_4$ sample produced from a large-volume crystal grown in the optimized Czohralski-based furnance in the framework of the CLYMENE project on development of Li$_2$MoO$_4$ crystals for next-generation rare decay search experiments. To detect scintillation of each crystal, we accompanied them with bolometric Ge light detectors; the design of the device coupled to the Li$_2$MoO$_4$ crystal allows a signal amplification based on the Neganov-Trofimov-Luke effect.

The $^{116}$CdWO$_4$ and Li$_2$MoO$_4$ thermal detectors show high performance, in particular the energy resolution of both detectors was measured as $\sim$2 keV FWHM for the baseline noise and $\sim$6 keV FWHM for the 2.6 MeV $\gamma$ quanta. The resolution of the  $^{116}$CdWO$_4$ bolometer is the best ever reported for large CdWO$_4$ detectors (with a volume about $\sim$30 cm$^3$). The resolution achieved with the Li$_2$MoO$_4$  bolometer is among the best achieved with thermal detectors containing this absorber material, particularly tested in the CROSS set-up. An ordinary bolometric Ge light detector, coupled to the efficient scintillator, was characterized with a baseline noise resolution around 100 eV FWHM (a factor 2 better than in the set-up before the suspension upgrade), while the performance of the Neganov-Trofimov-Luke device was improved to 16 eV FWHM noise resolution by applying a 60 V bias on the Al electrode. Despite of a clear improvement of noise conditions in the upgraded set-up, we still see pulse-tube induced vibration in the noise data of the detectors, which indicates on a noise pick-up by the cryostat wiring. 
A highly efficient scintillation-assisted particle identification has been achieved with both studied scintillating bolometers. By analysis of $\alpha$ events detected by the Li$_2$MoO$_4$ detector, we observed a clear surface contamination (introduced by the crystal polishing) together with traces of $^{210}$Po and $^{226}$Ra with a similar activity of $\sim$0.1 mBq/kg, while the $^{228}$Th activity is expected to be at least an order of magnitude lower. From the study of the $\gamma$($\beta$) spectrum acquired by the Li$_2$MoO$_4$ bolometer we found no clear evidence of the crystal bulk contamination by $^{40}$K and set an upper limit on its activity on the level of $\sim$6 mBq/kg. Thus, the Li$_2$MoO$_4$ crystal exhibits reasonably low level of radioactive contaminations despite the absence of additional purification of the starting material, typically done for rare-event search experiments.

\acknowledgments

This work is supported by the European Commission (Project CROSS, Grant No. ERC-2016-ADG, ID 742345) and by the Agence Nationale de la Recherche (Project CLYMENE; ANR-16-CE08-0018; Project CUPID-1; ANR-21-CE31-0014, ANR France). We acknowledge also the support of the P2IO LabEx (ANR-10-LABX0038) in the framework ``Investissements d'Avenir'' (ANR-11-IDEX-0003-01 -- Project ``BSM-nu'') managed by ANR, France. 
This work was also supported by the National Research Foundation of Ukraine under Grant No. 2020.02/0011 and by the National Academy of Sciences of Ukraine in the framework of the project ``Development of bolometric experiments for the search for double beta decay'', the grant number 0121U111684. 
Russian and Ukrainian scientists have given and give crucial contributions to CROSS. For this reason, the CROSS collaboration is particularly sensitive to the current situation in Ukraine. The position of the collaboration leadership on this matter, approved by majority, is expressed at \href{https://a2c.ijclab.in2p3.fr/en/a2c-home-en/assd-home-en/assd-cross/}{https://a2c.ijclab.in2p3.fr/en/a2c-home-en/assd-home-en/assd-cross/}.



\end{document}